\documentclass[showkeys,superscriptaddress,floatfix,prd,10pt,aps]{revtex4-2}
\usepackage{graphicx,epstopdf}
\usepackage[caption=false]{subfig}
\usepackage{appendix}
\usepackage[T1]{fontenc}
\usepackage{lmodern}
\usepackage[dvipsnames,x11names]{xcolor}
\usepackage[colorlinks=true,linkcolor=ForestGreen,citecolor=ForestGreen,urlcolor=NavyBlue]{hyperref}
\usepackage[sort&compress]{natbib}
\usepackage{lipsum}
\usepackage{morefloats}
\usepackage[pdf]{pstricks}
\usepackage{amsmath}
\usepackage{amssymb}
\usepackage{amsfonts}
\usepackage{rotating}
\usepackage{cancel}
\usepackage{mathtools}
\usepackage{bbm}
\usepackage{dsfont}
\usepackage{bbold}
\usepackage{multirow}
\usepackage{ulem}
\usepackage{physics}
\usepackage{orcidlink}
\usepackage{colortbl}

\def\xslash{x\!\!\!\slash }

\def\vel{\left|}
\def\ver{\right|}

\begin{document}

\title{%Unveiling the structure of axial-vector bottom-charm tetraquarks by their magnetic moments
Unveiling the underlying structure of axial-vector bottom-charm tetraquarks in the light of their magnetic moments 
}

\author{Ula\c{s}~\"{O}zdem\orcidlink{0000-0002-1907-2894}}%
\email[]{ulasozdem@aydin.edu.tr }
\affiliation{ Health Services Vocational School of Higher Education, Istanbul Aydin University, Sefakoy-Kucukcekmece, 34295 Istanbul, T\"{u}rkiye}

\date{\today}
 
\begin{abstract}
The magnetic moment yields an excellent framework to explore the inner structure of particles determined by the quark-gluon dynamics of QCD, as it is the leading-order response of a bound system to a weak external magnetic field. Motivated by this, in this study, the magnetic moments of possible axial-vector $T_{bc\bar u \bar u}$, $T_{bc\bar d \bar d}$, and  $T_{bc\bar u \bar d}$ tetraquarks are obtained with the help of light-cone QCD sum rules. For this purpose, we assume that these states are represented as a diquark-antidiquark picture with different structures and interpolating currents. The magnetic moment results derived using different diquark-antidiquark configurations differ substantially from each other. This can be translated into more than one tetraquark state with the same quantum number and quark content yet possessing different magnetic moments.  From the numerical results obtained, we have concluded that the magnetic moments of the $T_{bc}$ states can project their inner structure, which can be used for their quantum numbers and quark-gluon organization. The contribution of individual quarks to the magnetic moments is also analyzed for completeness. We hope that our predictions of the magnetic moments of the $T_{bc}$ tetraquarks, together with the results of other theoretical investigations of the spectroscopic parameters and decay widths of these interesting tetraquarks, may be valuable in the search for these states in future experiments and in unraveling the internal structure of these tetraquarks.
\end{abstract}

\maketitle

\section{motivation}\label{motivation}

Although the concept of hadrons with more complex structures than standard hadrons ($q \bar q /qqq$) has been known for some time, it became experimentally proven in 2003 when the Belle Collaboration observed the X(3872) particle~\cite{Belle:2003nnu}. Since then, various non-conventional hadron candidates have been observed by different collaborations. Most observed unconventional hadron candidates contain hidden-heavy quarks until 2021. In 2021, the LHCb collaboration observed the $T_{cc}(3875)^+$ particle \cite{LHCb:2021vvq,LHCb:2021auc}, which is the first candidate for a doubly-heavy tetraquark with [cc]  quarks. The discovery of unconventional hadron states, including hidden-heavy and doubly-heavy states, is a cornerstone of hadron physics and is an active field of study both experimentally and theoretically. Recent developments on unconventional states, as well as experimental and theoretical studies, are presented in Refs.~\cite{Esposito:2014rxa,Esposito:2016noz,Olsen:2017bmm,Lebed:2016hpi,Nielsen:2009uh,Brambilla:2019esw,Agaev:2020zad,Chen:2016qju,Ali:2017jda,Guo:2017jvc,Liu:2019zoy,Yang:2020atz,Dong:2021juy,Dong:2021bvy,Meng:2022ozq, Chen:2022asf}, and references therein. It's important to note, however, that despite the numerous studies that have been conducted, the internal structure of these unconventional states is still incompletely understood. It remains one of the most important questions to be answered.

  Another class of tetraquark states, those containing the heavy diquarks bc, is also being studied by physicists. One of the main reasons for this is that in these situations tetraquarks built from bc may be stable against the strong and electromagnetic decays. Therefore, the states with different quantum numbers composed of heavy bc diquarks have also been intensively studied~
\cite{Padmanath:2023rdu,Kim:2022mpa,Deng:2021gnb,Agaev:2020mqq,Cheng:2020wxa,Lu:2020rog,Wang:2020jgb,Agaev:2020dba,Agaev:2019lwh,Sundu:2019feu,Carames:2018tpe,Deng:2018kly,Agaev:2018khe,Chen:2013aba,Mutuk:2023oyz,Ma:2023int}. The study of the properties of $T_{bc}$ tetraquarks along with the doubly-bottom and doubly-charm states could result in a deeper understanding of the strong interaction dynamics over a broad regime of quark masses, spanning from charm to bottom quarks. Further investigations from various theoretical perspectives are necessary to provide a comprehensive understanding of the properties of the $T_{bc}$ tetraquarks, which should be useful in future experimental studies. 

To understand the internal structure and geometric shape of $T_{bc}$ tetraquarks, as well as the complex nature of QCD, it may be useful to study their properties, such as electromagnetic or radiative transitions.  
Motivated by those reasons, in this study, we use the QCD light-cone sum rules (LCSR) method to extract the magnetic moment of $T_{bc}$ tetraquarks with spin-parity quantum numbers $J^P = 1^+$. In doing so, we assume that these states are represented as a diquark-antidiquark picture with different structures and interpolating currents. 
Moreover, the magnetic moments of vector $T_{bc}$ tetraquarks spin-parity quantum numbers $J^P = 1^-$ can also be calculated. However, spectroscopic analyses have demonstrated that these states are not stable and cannot form bound states i.e., their masses lies above the related two-meson thresholds (see Refs.~\cite{Kim:2022mpa,Deng:2021gnb,Cheng:2020wxa,Ma:2023int}). Consequently, the magnetic moments of these vector $T_{bc}$ tetraquarks are not analyzed. As analytical sensitivity and the inclusion of next-to-leading-order (NLO) contributions improve in the future, the magnetic moment analyses of these states can be re-evaluated based on the results obtained. For the time being, however, we do not intend to perform these analyses. 
For the aforementioned reasons, this study will pay attention to  the magnetic moments of the axial-vector $T_{bc}$ states by means of the LCSR. 
The LCSR method provides highly effective and predictable results and is a powerful non-perturbative method for studying the dynamic and static properties of conventional and unconventional hadrons. The LCSR method calculates the correlation function, which is the crucial part of the method, in two kinematic regions: in terms of hadrons and in terms of QCD parameters using the Operator Product Expansion (OPE) and the photon's distribution amplitudes.  By relating the results of these two representations, via dispersion integrals and by using the quark-hadron duality ansatz, one can obtain the magnetic moments associated with other parameters. Borel transformations and continuum subtraction techniques are then carried out to suppress the contributions of the higher states and the continuum. The LCSR for the considered problem is derived by matching the coefficients of the Lorentz structures used in the analysis~\cite{Chernyak:1990ag, Braun:1988qv, Balitsky:1989ry}. The use of LCSR to obtain electromagnetic properties for doubly-heavy baryons, tetraquarks and pentaquarks  is illustrated in Refs.~\cite{Ozdem:2024lpk,Ozdem:2023cri,Azizi:2023gzv,Ozdem:2022yhi,Ozdem:2022vip,Ozdem:2021hmk,Azizi:2021aib,Ozdem:2019zis,Ozdem:2018uue}.

 The article is structured in the following way: In Sec. \ref{formalism}, we consider the $T_{bc}$ tetraquarks in the diquark-antidiquark context and perform the LCSR method to extract their magnetic moments. For the magnetic moment of the $T_{bc}$ tetraquarks, the numerical analysis and discussion are presented in Sec. \ref{numerical}.  The final section summarizes and discusses the results.

 \begin{widetext}
 
\section{Theoretical framework}\label{formalism}

The magnetic moment of the $T_{bc}$ tetraquarks can be derived using the LCSR method by utilizing the following correlation function: 

\begin{equation}
 \label{edmn00}
\Pi _{\mu \nu \alpha }(p,q)=i^2\int d^{4}x\,\int d^{4}y\,e^{ip\cdot x+iq \cdot y}\,
\langle 0|\mathcal{T}\{J_{\mu}(x) J_{\alpha}(y)
J_{\nu }^{\dagger }(0)\}|0\rangle,  
\end{equation}%
where $J_\alpha$ is the electromagnetic current, $J_{\mu(\nu)}(x)$ is the interpolating currents for the states under consideration with the quantum numbers $ J^{P} = 1^{+}$.
The relevant currents are written as below
\begin{align}
 J_\alpha &=\sum_{q= u,d,c,b} e_q \bar q \gamma_\alpha q,\\
 %\nonumber\\
J_{\mu }^{1}(x) &= \big[b^{a^T} (x) C \gamma_5 c^b (x)\big]\big[\bar q_1^a (x) \gamma_\mu C \bar q_2^{b^T} (x) - \bar q_1^b (x) \gamma_\mu C \bar q_2^{a^T} (x)\big],\\
%\nonumber\\
%%%%%%%%%%%%%%%%
J_{\mu }^{2}(x) &= \big[b^{a^T} (x) C \gamma_\mu c^b (x)\big]\big[\bar q_1^a (x) \gamma_5C \bar q_2^{b^T} (x) - \bar q_1^b (x) \gamma_5 C \bar q_2^{a^T} (x)\big],
\label{curr}
\end{align}
where $e_q$ is the electric charge of the corresponding quark, $a$ and $b$ are color indices, $C$ is the charge conjugation operator and;  $q_1(x)$ and $q_2(x)$ denote the $u(x)$ or  $d(x)$  quarks. As noted above, since the interpolating currents $J_{\mu }^{1}(x)$ and $J_{\mu }^{2}(x)$ have the same quantum numbers, they would most likely couple to the same tetraquark states.

Rewriting the correlation function by means of the external background electromagnetic field is more practical from a technical point of view,
\begin{equation}
 \label{edmn01}
\Pi _{\mu \nu }(p,q)=i\int d^{4}x\,e^{ip\cdot x}\langle 0|\mathcal{T}\{J_{\mu}(x)
J_{\nu }^{\dagger }(0)\}|0\rangle_{F}, 
\end{equation}%
where  F is the external background electromagnetic field and 
$F_{\alpha\beta}= i (\varepsilon_\alpha q_\beta-\varepsilon_\beta q_\alpha) e^{-iq\cdot x}$ with
$q_\alpha$ and $\varepsilon_\beta$  being the four-momentum and polarization
of the external background electromagnetic field. The external background electromagnetic field approach has a main advantage in that it explicitly separates the soft and hard photon emissions in a gauge-invariant way~\cite{Ball:2002ps}.
As is well known, the external background electromagnetic field is made arbitrarily small, and in this case the correlation function in Eq. (\ref{edmn01}) could be expanded in powers of the external background electromagnetic field and rewritten as follows, 
 \begin{equation}
\Pi _{\mu \nu }(p,q) = \Pi _{\mu \nu }^{(0)}(p,q) + \Pi _{\mu \nu }^{(1)}(p,q)+.... ,
\end{equation}
where $\Pi _{\mu \nu }^{(0)}(p,q)$ is the correlation function in the lack of an external background electromagnetic field and corresponds to the mass sum rules, which is not related to our context and  $\Pi _{\mu \nu }^{(1)}(p,q)$ corresponds to the single photon emission~\cite{Ball:2002ps,Novikov:1983gd,Ioffe:1983ju}. 
As a result, in order to obtain the electromagnetic properties of the hadrons within the LCSR, all we have to do is calculate the $\Pi _{\mu \nu }^{(1)}(p,q)$ term.

%%%%%%%%%%%%%%%%%%%%%%%%%%%%%%%%%%%%

Having made these clarifications, we may proceed to obtain the LCSR for the magnetic moment of the $T_{bc}$ states.  The starting point of our analysis will be the calculation of the hadronic representation of the correlation function. The correlation function between the interpolating currents is calculated by inserting a complete set of hadronic states that carry the same quantum numbers as the interpolating current and integrating over four-x.  With these operations, the result is achieved as
\begin{align}
\label{edmn04}
\Pi_{\mu\nu}^{Had} (p,q) &= {\frac{\langle 0 \mid J_\mu (x) \mid
T_{bc}(p) \rangle}{p^2 - m_{T_{bc}}^2}} \langle T_{bc} (p) \mid T_{bc} (p+q) \rangle_F %\nonumber\\
%& \times
\frac{\langle T_{bc} (p+q) \mid J_{\nu }^{\dagger } (0) \mid 0 \rangle}{(p+q)^2 - m_{T_{bc}}^2} + \mbox{higher states},
\end{align}
%where higher states are shown by dots.   
The expression $\langle 0 \mid J_\mu (x) \mid T_{bc}(p) \rangle$ is written with respect to several hadron measurables as follows
\begin{align}
\label{edmn05}
\langle 0 \mid J_\mu (x) \mid T_{bc} (p) \rangle = m_{T_{bc}} f_{T_{bc}} \varepsilon_\mu^\theta\,,
\end{align}
with  $f_{T_{bc}}$ and $ \varepsilon_\mu^\theta\ $   are the current coupling and the polarization vector of particles under consideration, respectively.  
The radiative matrix element in Eq. (\ref{edmn04}) is  described in the following way~\cite{Brodsky:1992px}:
\begin{align}
\label{edmn06}
\langle T_{bc}(p,\varepsilon^\theta) \mid  T_{bc} (p+q,\varepsilon^{\delta})\rangle_F &= - \varepsilon^\tau (\varepsilon^{\theta})^\alpha (\varepsilon^{\delta})^\beta 
%\nonumber\\
%&
\Bigg[ G_1(Q^2)~ (2p+q)_\tau ~g_{\alpha\beta}  
%\nonumber\\
%&
+ G_2(Q^2)~ ( g_{\tau\beta}~ q_\alpha -  g_{\tau\alpha}~ q_\beta)
\nonumber\\ 
&
- \frac{1}{2 m_{T_{bc}}^2} G_3(Q^2)~ (2p+q)_\tau 
%\nonumber\\
%&\times ~
q_\alpha q_\beta  \Bigg],
\end{align}
where %the $\varepsilon^\tau$ is the polarization vector of the photon.  Here, 
$G_i(Q^2)$'s are invariant form factors,  with  $Q^2=-q^2$.  
Utilizing Eqs.~(\ref{edmn04})-(\ref{edmn06}), the correlation function can be extracted in the following way:
%\begin{widetext}
%
\begin{align}
\label{edmn09}
 \Pi_{\mu\nu}^{Had}(p,q) &=  \frac{\varepsilon_\rho \, m_{T_{bc}}^2 f_{T_{bc}}^2}{ [m_{T_{bc}}^2 - (p+q)^2][m_{T_{bc}}^2 - p^2]}
 %\nonumber\\
 %&
 \bigg\{G_1(Q^2)(2p+q)_\rho\bigg[g_{\mu\nu}-\frac{p_\mu p_\nu}{m_{T_{bc}}^2}
 %\nonumber\\
 %&
 -\frac{(p+q)_\mu (p+q)_\nu}{m_{T_{bc}}^2}+\frac{(p+q)_\mu p_\nu}{2m_{T_{bc}}^4}\nonumber\\
 & \times (Q^2+2m_{T_{bc}}^2)
 \bigg]
 + G_2 (Q^2) \bigg[q_\mu g_{\rho\nu}  
 %\nonumber\\
 %&
 - q_\nu g_{\rho\mu}-
\frac{p_\nu}{m_{T_{bc}}^2}  \big(q_\mu p_\rho - \frac{1}{2}
Q^2 g_{\mu\rho}\big) 
%\nonumber\\
%&
+
\frac{(p+q)_\mu}{m_{T_{bc}}^2}  \big(q_\nu (p+q)_\rho+ \frac{1}{2}
Q^2 g_{\nu\rho}\big) 
\nonumber\\
&-  
\frac{(p+q)_\mu p_\nu p_\rho}{m_{T_{bc}}^4} \, Q^2
\bigg]
%\nonumber\\
%&
-\frac{G_3(Q^2)}{m_{T_{bc}}^2}(2p+q)_\rho \bigg[
q_\mu q_\nu -\frac{p_\mu q_\nu}{2 m_{T_{bc}}^2} Q^2 
%\nonumber\\
%&
+\frac{(p+q)_\mu q_\nu}{2 m_{T_{bc}}^2} Q^2
-\frac{(p+q)_\mu q_\nu}{4 m_{T_{bc}}^4} Q^4\bigg]
\bigg\}\,.
\end{align}
%\end{widetext}
The magnetic form factor ($F_M(Q^2)$) is expressed as the $G_2(Q^2)$ form factor:  
\begin{align}
\label{edmn07}
&F_M(Q^2) = G_2(Q^2)\,. 
\end{align}
%where $\tau=Q^2/4 m_{T_{bc}}^2$ with $Q^2=-q^2$. 
In the static limit $(Q^2=0)$, the form factor $F_M(0)$ is associated with the magnetic moment ($\mu$) by the following equation
\begin{align}
\label{edmn08}
& \mu  = \frac{e}{2 m_{T_{bc}}}\,F_M(0).
\end{align}
For the physical quantity under investigation, a representation of the analysis in terms of hadronic parameters is obtained. Now it is time to get a representation of the analysis in terms of the quark-gluon parameters.

The QCD part of the correlation function is calculated in the 
deep Euclidean domain by means of OPE. After performing simple calculations, we get the following expressions for the $T_{bc}$ states
%\begin{widetext}
\begin{align}
\Pi _{\mu \nu }^{\mathrm{QCD}-J_\mu^1}(p,q)&=i\int d^{4}xe^{ip\cdot x} \langle 0 \mid \bigg\{  \mathrm{Tr}%
\Big[ \gamma _{5 }S_{c}^{b^{\prime }b}(x)\gamma _{5}S_{b}^{aa^{\prime }}(x)\Big]    \mathrm{Tr}\Big[ \gamma _{\mu} \widetilde{S}_{q_2}^{b^{\prime}b}(-x)\gamma _{\nu} S_{q_1}^{a^{\prime }a}(-x)\Big] \nonumber\\
%%%%%%%%%%%%%%%%%%%%%%%%%%%%%%%%%%%%%%%%%%%%%%%%%%%%%%%%%%%%%%%%%%%%%%%%%%%%%%%%%%%%
&-\mathrm{Tr}%
\Big[ \gamma _{5 }S_{c}^{b^{\prime }b}(x)\gamma _{5}S_{b}^{aa^{\prime }}(x)\Big]    \mathrm{Tr}\Big[ \gamma _{\mu} \widetilde{S}_{q_1 q_2}^{a^{\prime}b}(-x)\gamma _{\nu} S_{q_2q_1}^{b^{\prime }a}(-x)\Big] \nonumber\\ 
%%%%%%%%%%%%%%%%%%%%%%%%%%%%%%%%%%%%%%%%%%%%%%%%%%%%%%%%%%%%%%%%%%%%%%%%%%%%%%%%%%%%
&-\mathrm{Tr}%
\Big[ \gamma _{5}S_{c}^{b^{\prime }b}(x)\gamma _{5}S_{b}^{aa^{\prime }}(x)\Big]    \mathrm{Tr}\Big[ \gamma _{\mu} \widetilde{S}_{q_{2}}^{b^{\prime}a}(-x)\gamma _{\nu} S_{q_{1}}^{a^{\prime }b}(-x)\Big] \nonumber\\ 
%%%%%%%%%%%%%%%%%%%%%%%%%%%%%%%%%%%%%%%%%%%%%%%%%%%%%%%%%%%%%%%%%%%
& +\mathrm{Tr}%
\Big[ \gamma _{5 }S_{c}^{b^{\prime }b}(x)\gamma _{5}S_{b}^{aa^{\prime }}(x)\Big]    \mathrm{Tr}\Big[ \gamma _{\mu} \widetilde{S}_{q_2q_1}^{a^{\prime}a}(-x)\gamma _{\nu} S_{q_1 q_2}^{b^{\prime }b}(-x)\Big] \nonumber\\
%%%%%%%%%%%%%%%%%%%%%%%%%%%%%%%%%%%%%%%%%%%%%%%%%%%%%%%%%%%%%%%%%%%%%%%%%%%%%%%%%%%%
&-\mathrm{Tr}%
\Big[ \gamma _{5}S_{c}^{b^{\prime }b}(x)\gamma _{5}S_{b}^{aa^{\prime }}(x)\Big]    \mathrm{Tr}\Big[ \gamma _{\mu} \widetilde{S}_{q_{2}}^{a^{\prime}b}(-x)\gamma _{\nu} S_{q_{1}}^{b^{\prime }a}(-x)\Big] \nonumber\\ 
%%%%%%%%%%%%%%%%%%%%%%%%%%%%%%%%%%%%%%%%%%%%%%%%%%%%%%%%%%%%%%%%%%%%%%%%%%%%%%%%%%%%%%%%%%%%%%
&+\mathrm{Tr}%
\Big[ \gamma _{5}S_{c}^{b^{\prime }b}(x)\gamma _{5}S_{b}^{aa^{\prime }}(x)\Big]    \mathrm{Tr}\Big[ \gamma _{\mu} \widetilde{S}_{q_2q_1}^{b^{\prime}b}(-x)\gamma _{\nu} S_{q_1 q_2}^{a^{\prime }a}(-x)\Big] \nonumber\\ 
%%%%%%%%%%%%%%%%%%%%%%%%%%%%%%%%%%%%%%%%%%%%%%%%%%%%%%%%%%%%%%%%%%%%%%%%%%%%%%%%%%%%%%%%%%%%
&+\mathrm{Tr}%
\Big[ \gamma _{5}S_{c}^{b^{\prime }b}(x)\gamma _{5}S_{b}^{aa^{\prime }}(x)\Big]    \mathrm{Tr}\Big[ \gamma _{\mu} \widetilde{S}_{q_{2}}^{a^{\prime}a}(-x)\gamma _{\nu} S_{q_{1}}^{b^{\prime }b}(-x)\Big] \nonumber\\ 
%%%%%%%%%%%%%%%%%%%%%%%%%%%%%%%%%%%%%%%%%%%%%%%%%%%%%%%%%%%%%%%%%%%%%%%%%%%
&-\mathrm{Tr}%
\Big[ \gamma _{5}S_{c}^{b^{\prime }b}(x)\gamma _{5}S_{b}^{aa^{\prime }}(x)\Big]    
\mathrm{Tr}\Big[ \gamma _{\mu} \widetilde{S}_{q_2q_1}^{b^{\prime}a}(-x)\gamma _{\nu} S_{q_1 q_2}^{a^{\prime }b}(-x)\Big]  
\bigg\} \mid 0 \rangle_{F} ,  \label{eq:QCDSide}
\end{align}%
\begin{align}
\Pi _{\mu \nu }^{\mathrm{QCD}-J_\mu^2}(p,q)&=i\int d^{4}xe^{ip\cdot x} \langle 0 \mid \bigg\{  \mathrm{Tr}%
\Big[ \gamma _{\mu }S_{c}^{b^{\prime }b}(x)\gamma _{\nu}S_{b}^{aa^{\prime }}(x)\Big]    \mathrm{Tr}\Big[ \gamma _{5} \widetilde{S}_{q_2}^{b^{\prime}b}(-x)\gamma _{5} S_{q_1}^{a^{\prime }a}(-x)\Big] \nonumber\\
%%%%%%%%%%%%%%%%%%%%%%%%%%%%%%%%%%%%%%%%%%%%%%%%%%%%%%%%%%%%%%%%%%%%%%%%%%%%%%%%%%%%
&-\mathrm{Tr}%
\Big[ \gamma _{\mu }S_{c}^{b^{\prime }b}(x)\gamma _{\nu}S_{b}^{aa^{\prime }}(x)\Big]    \mathrm{Tr}\Big[ \gamma _{5} \widetilde{S}_{q_1 q_2}^{a^{\prime}b}(-x)\gamma _{5} S_{q_2q_1}^{b^{\prime }a}(-x)\Big] \nonumber\\ 
%%%%%%%%%%%%%%%%%%%%%%%%%%%%%%%%%%%%%%%%%%%%%%%%%%%%%%%%%%%%%%%%%%%%%%%%%%%%%%%%%%%%
&-\mathrm{Tr}%
\Big[ \gamma _{\mu }S_{c}^{b^{\prime }b}(x)\gamma _{\nu}S_{b}^{aa^{\prime }}(x)\Big]    \mathrm{Tr}\Big[ \gamma _{5} \widetilde{S}_{q_{2}}^{b^{\prime}a}(-x)\gamma _{5} S_{q_{1}}^{a^{\prime }b}(-x)\Big] \nonumber\\ 
%%%%%%%%%%%%%%%%%%%%%%%%%%%%%%%%%%%%%%%%%%%%%%%%%%%%%%%%%%%%%%%%%%%
& +\mathrm{Tr}%
\Big[ \gamma _{\mu }S_{c}^{b^{\prime }b}(x)\gamma _{\nu}S_{b}^{aa^{\prime }}(x)\Big]    \mathrm{Tr}\Big[ \gamma _{5} \widetilde{S}_{q_2q_1}^{a^{\prime}a}(-x)\gamma _{5} S_{q_1 q_2}^{b^{\prime }b}(-x)\Big] \nonumber\\
%%%%%%%%%%%%%%%%%%%%%%%%%%%%%%%%%%%%%%%%%%%%%%%%%%%%%%%%%%%%%%%%%%%%%%%%%%%%%%%%%%%%
&-\mathrm{Tr}%
\Big[ \gamma _{\mu }S_{c}^{b^{\prime }b}(x)\gamma _{\nu}S_{b}^{aa^{\prime }}(x)\Big]    \mathrm{Tr}\Big[ \gamma _{5} \widetilde{S}_{q_{2}}^{a^{\prime}b}(-x)\gamma _{5} S_{q_{1}}^{b^{\prime }a}(-x)\Big] \nonumber\\ 
%%%%%%%%%%%%%%%%%%%%%%%%%%%%%%%%%%%%%%%%%%%%%%%%%%%%%%%%%%%%%%%%%%%%%%%%%%%%%%%%%%%%%%%%%%%%%%
&+\mathrm{Tr}%
\Big[ \gamma _{\mu }S_{c}^{b^{\prime }b}(x)\gamma _{\nu}S_{b}^{aa^{\prime }}(x)\Big]    \mathrm{Tr}\Big[ \gamma _{5} \widetilde{S}_{q_2q_1}^{b^{\prime}b}(-x)\gamma _{5} S_{q_1 q_2}^{a^{\prime }a}(-x)\Big] \nonumber\\ 
%%%%%%%%%%%%%%%%%%%%%%%%%%%%%%%%%%%%%%%%%%%%%%%%%%%%%%%%%%%%%%%%%%%%%%%%%%%%%%%%%%%%%%%%%%%%
&+\mathrm{Tr}%
\Big[ \gamma _{\mu }S_{c}^{b^{\prime }b}(x)\gamma _{\nu}S_{b}^{aa^{\prime }}(x)\Big]    \mathrm{Tr}\Big[ \gamma _{5} \widetilde{S}_{q_{2}}^{a^{\prime}a}(-x)\gamma _{5} S_{q_{1}}^{b^{\prime }b}(-x)\Big] \nonumber\\ 
%%%%%%%%%%%%%%%%%%%%%%%%%%%%%%%%%%%%%%%%%%%%%%%%%%%%%%%%%%%%%%%%%%%%%%%%%%%
&-\mathrm{Tr}%
\Big[ \gamma _{\mu }S_{c}^{b^{\prime }b}(x)\gamma _{\nu}S_{b}^{aa^{\prime }}(x)\Big]    \mathrm{Tr}\Big[ \gamma _{5} \widetilde{S}_{q_2q_1}^{b^{\prime}a}(-x)\gamma _{5} S_{q_1 q_2}^{a^{\prime }b}(-x)\Big]  
\bigg\} \mid 0 \rangle_{F} ,  \label{eq:QCDSide2}
\end{align}%
%\end{widetext}
where 
$\widetilde{S}_{Q(q)}^{ij}(x)=CS_{Q(q)}^{ij\rm{T}}(x)C$. $S_{q_{i}q_{j}}(x)$ exists when $q_i=q_j$ but it vanishes when $q_i\neq q_j$.  The $S_{Q}(x)$ and $S_{q}(x)$ are the full propagators of heavy and light quarks, and they have the following forms~\cite{Yang:1993bp, Belyaev:1985wza},
\begin{align}
\label{edmn13}
S_{q}(x)&= S_q^{free}(x) 
- \frac{\langle \bar qq \rangle }{12} \Big(1-i\frac{m_{q} \xslash}{4}   \Big)
%\nonumber\\
%&
- \frac{ \langle \bar qq \rangle }{192}
m_0^2 x^2  \Big(1 %\nonumber\\
%&
  -i\frac{m_{q} \xslash}{6}   \Big)
+\frac {i g_s~G^{\mu \nu} (x)}{32 \pi^2 x^2} 
%\nonumber\\
%& \times 
\Bigg[\rlap/{x} 
\sigma_{\mu \nu} +  \sigma_{\mu \nu} \rlap/{x}
 \Bigg],\\
%\nonumber\\
%\end{align}%
%and
%
%\begin{align}
S_{Q}(x)&=S_Q^{free}(x)
%\nonumber\\
%&
-\frac{m_{Q}\,g_{s}\, G^{\mu \nu}(x)}{32\pi ^{2}} \Bigg[ (\sigma _{\mu \nu }{\xslash}
 % \nonumber\\
%&
+{\xslash}\sigma _{\mu \nu }) 
    \frac{K_{1}\big( m_{Q}\sqrt{-x^{2}}\big) }{\sqrt{-x^{2}}}
   %\nonumber\\
  %&
 +2\sigma_{\mu \nu }K_{0}\big( m_{Q}\sqrt{-x^{2}}\big)\Bigg],
 \label{edmn14}
\end{align}%
with  
\begin{align}
 S_q^{free}(x)&=\frac{1}{2 \pi x^2}\Big(i \frac{\xslash}{x^2}- \frac{m_q}{2}\Big),\\
 \nonumber\\
 S_Q^{free}(x)&=\frac{m_{Q}^{2}}{4 \pi^{2}} \Bigg[ \frac{K_{1}\big(m_{Q}\sqrt{-x^{2}}\big) }{\sqrt{-x^{2}}}
+i\frac{{\xslash}~K_{2}\big( m_{Q}\sqrt{-x^{2}}\big)}
{(\sqrt{-x^{2}})^{2}}\Bigg],
\end{align}
where $m_0$ is defined through the quark-gluon mixed condensate $ m_0^2= \langle 0 \mid \bar  q\, g_s\, \sigma_{\mu\nu}\, G^{\mu\nu}\, q \mid 0 \rangle / \langle \bar qq \rangle $, $G^{\mu\nu}$ is the gluon field-strength tensor, and $K_i$'s being the modified second type Bessel functions.  

Eqs. (\ref{eq:QCDSide}) and (\ref{eq:QCDSide2}) contain two different contributions, perturbative and non-perturbative, i.e. the photon interacts with the quarks perturbatively or non-perturbatively, that need to be included in the analysis. We will briefly describe how these contributions are included in the analysis, as their calculation is lengthy and tedious. To see how these contributions are calculated, we refer the reader to the Refs.~\cite{Ozdem:2022eds,Ozdem:2022vip}. 

To determine the perturbative contribution, it is sufficient to perform the following substitution
\begin{align}
\label{free}
S^{free}(x) \longrightarrow \int d^4z\, S^{free} (x-z)\,\rlap/{\!A}(z)\, S^{free} (z)\,.
\end{align}
To determine the perturbative contribution, the following formulas are to be employed
 \begin{align}
\label{edmn21}
S_{\alpha\beta}^{ab}(x) \longrightarrow -\frac{1}{4} \big[\bar{q}^a(x) \Gamma_i q^b(0)\big]\big(\Gamma_i\big)_{\alpha\beta},
\end{align}
where $\Gamma_i = \{\textbf{1}$, $\gamma_5$, $\gamma_\mu$, $i\gamma_5 \gamma_\mu$, $\sigma_{\mu\nu}/2\}$. When calculating the perturbative contribution, one propagator is used in the equation, and the remaining propagators are analyzed using the free part. For the non-perturbative contribution, we include all remaining propagators in the analysis as full propagators. When non-perturbative contributions are included in the analysis, matrix elements like $\langle \gamma(q)\vel \bar{q}(x) \Gamma_i G_{\alpha\beta}q(0) \ver 0\rangle$ and $\langle \gamma(q)\vel \bar{q}(x) \Gamma_i q(0) \ver 0\rangle$ appear. These matrix elements, which are written associated with the photon wave functions, have an important role in the extraction of non-perturbative contributions (see Ref.~\cite{Ball:2002ps} for details on photon distribution amplitudes (DAs)). As a result of the above procedures, the correlation function in QCD representation has been obtained through the quark-gluon properties using photo DAs.  By integrating over x, the correlation function in the momentum representation has been derived in a simple way. 

Utilizing dispersion relations that consider the coefficients of the same Lorentz structures, i.e., $(\varepsilon_\mu q_\nu- \varepsilon_\nu q_\mu)$, the results obtained by performing calculations on both sides of the correlation function are compared. 
In the last step, we carry out Borel transformation on the variables $-p^2$ and $-(p+q)^2$ to dominate contributions from the continuum and the higher states and boost ground states to get 
\begin{equation}
    \begin{aligned}
                \mu_{T_{bc}} m^2_{T_{bc}} f^2_{T_{bc}}e^{-\frac{m_2^2}{M_1^2}}e^{-\frac{m_1^2}{M_2^2}} &= \int_0^\infty ds_1 \int_0^\infty ds_2 \, e^{-\frac{s_1}{M_1^2}-\frac{s_2}{M_2^2}}\rho(s_1,s_2).\\
    \end{aligned}
    \label{correlated}
\end{equation}
 To acquire the magnetic moment within the LCSR, the contributions from the higher states and the continuum have been extracted utilizing quark-hadron duality ansatz: 
\begin{equation}
    \rho(s_1,s_2) \simeq \rho^{OPE}(s_1,s_2) ~~\mbox{if}~~ (s_1,s_2) \not\in {\mathbb D},
\end{equation}
where $\mathbb D$ is a domain in the $(s_1,s_2)$ plane. 
%Generally, the domain $\mathbb D$ is a rectangular region defined by $s_1<s_{10}$ and $s_2<s_{20}$ for some constants $s_{10}$ and $s_{20}$, or a triangular region. 
In the present study, for brevity, continuum subtraction is performed 
via selecting $\mathbb D$ as the region determined as $s \equiv s_1 u_0 + s_2 \bar u_0 < s_0$ where $u_0\equiv \frac{M_2^2}{M_1^2+M_2^2}$ and $\bar u_0 = 1 - u_0$. Defining  a second variable $u=\frac{s_1 u_0}{s}$, the integral in the $(s_1,s_2)$ plane can be defined as:
\begin{equation}
    \int_0^\infty ds_1 \int_0^\infty ds_2 \,e^{-\frac{s_1}{M_1^2}-\frac{s_2}{M_2^2}}\, \rho(s_1,s_2) = \int_0^\infty ds \,\rho(s)\, e^{-\frac{s}{M^2}},
\end{equation}
where 
\begin{equation}
{M^2}= \frac{M_1^2 M_2^2}{M_1^2+M_2^2}, ~\rm{and}  ~
    \rho(s) = \frac{s}{u_0 \bar u_0} \int_0^1 du \rho\left( s\frac{u}{u_0}, s \frac{\bar u}{\bar u_0} \right). 
\end{equation}
In the problem under review, the masses of the initial and final state tetraquarks are identical, hence we can set $M_1^2 = M_2^2 = 2M^2$, which yields $u_0=\frac{1}{2}$.

The procedures described above yield the following sum rules for the magnetic moment of the $T_{bc}$ states at the point $Q^2 = 0$, 
\begin{align}
\label{jmu1}
 &\mu_{T_{bc}}^{j_\mu^1}\,  = \frac{e^{\frac{m_{T_{bc}}^{2}}{\rm{M^2}}}}{ m^{2}_{T_{bc}} f^{2}_{T_{bc}}} \,\, \rho_1(\rm{M^2},\rm{s_0}),\\
% \nonumber\\
% \end{align}
% \begin{align}
 &\mu_{T_{bc}}^{j_\mu^2}\,  = \frac{e^{\frac{m_{T_{bc}}^{2}}{\rm{M^2}}}}{ m^{2}_{T_{bc}} f^{2}_{T_{bc}}} \,\,\rho_2(\rm{M^2},\rm{s_0}),
 \label{jmu2}
\end{align}
where the values of $m_{T_{bc}}$ and $f_{T_{bc}}$ given in Eq. (\ref{jmu1}) are the results obtained using $J_\mu^1$ current, while the values of $m_{T_{bc}}$ and $f_{T_{bc}}$ in Eq. (\ref{jmu2}) are the values obtained using $J_\mu^2$ current and their numerical values will be given in the next section.
%The expressions of the functions $\Delta_1^{QCD}(\rm{M^2},\rm{s_0})$ and $\Delta_2^{QCD}(\rm{M^2},\rm{s_0})$ are given by
%
In Eqs. (\ref{jmu1})-(\ref{jmu2}), the $\rho_1(\rm{M^2},\rm{s_0})$ and $\rho_2(\rm{M^2},\rm{s_0})$ are given by
\begin{align}
\rho_1(\rm{M^2},\rm{s_0})&= -\frac{(e_{q_1} + 2 e_{q_{12}} + e_{q_2})}{2 ^{20} \times 3^2 \times 5^2 \times 7  \pi^5}\Bigg[
28 m_b m_c \big(19 I[0, 5] + 65 I[1, 4]\big) - 9 \big(4 I[0, 6] + 129 I[1, 5]\big)\Bigg]\nonumber\\
&+\frac {m_c^2 m_{q_1}C_1 C_2} {2^{13} \times 3^3 \pi^3} (e_ {q_ 1} + 
   2 e_ {q_ {12}} + e_ {q_ 2}) I[0, 1]\, I_2[h_{\gamma}]\nonumber\\
   &
   +\frac {5 m_c^2 C_1 f_ {3\gamma}} {2^{19} \times 3^4 \pi^3} (e_ {q_ 1} - 7 e_ {q_ {12}} - 2 e_ {q_ 2}) I[0, 
  2] I_ 1[\mathcal V]\nonumber\\
  &-\frac {11 m_c^2 C_2} {2^{13} \times 3^3 \times 5 \pi^3} (e_ {q_ 1} m_ {q_ {2}} + 
   e_ {q_ {12}} m_ {q_ {12}} + 2 e_ {q_ 2} m_ {q_ {1}}) I[0, 
   3]\, I_2[h_ {\gamma}]\nonumber\\
   &
   +\frac {11 m_c^2 f_ {3\gamma}} {2^{17} \times 3^3 \times 5 \pi^3} (e_ {q_ 1} + e_ {q_ {12}} + e_ {q_ 2}) I[0, 
   4]\, I_ 1[\mathcal V],
\label{app1}
\end{align}
\begin{align}
\rho_2(\rm{M^2},\rm{s_0})&= -\frac {m_b m_c (m_ {q_ {1}} m_ {q_ {2}} + 
     m_ {q_ {12}} m_ {q_ {21}}) } {2 ^{16} \times 3^2  \pi^5} (4 e_b - 5 e_c)  I[0, 4]\nonumber\\
     &
     +\frac {1} {2 ^{19} \times 3^2 \times 5^2  \pi^5}\Bigg[ -e_b \big (9 m_ {q_ {1}} m_ {q_ {2}} + 
       9 m_ {q_ {12}} m_ {q_ {21}} + 128 m_b m_c\big) + 
    e_c \big (27 m_ {q_ {1}} m_ {q_ {2}} + 
        27 m_ {q_ {12}} m_ {q_ {21}} \nonumber\\
        &+ 160 m_b m_c\big) \Bigg] I[0, 5]\nonumber\\
        &
        -\frac {(3 e_b - 4 e_c) } {2 ^{19} \times 3^2 \times 5^2  \pi^5} I[0,6]\nonumber\\
     &
        -\frac {(7 e_b - 9 e_c) } {2 ^{19}  \times 5  \pi^5} (m_ {q_ {1}} m_ {q_ {2}} + 
   m_ {q_ {12}} m_ {q_ {21}}) I[1, 4]\nonumber\\
     &
   -\frac {(9 e_b - 11 e_c) } {2 ^{18}  \times 3 \times 5  \pi^5} I[1, 5]\nonumber\\
   &+\frac {m_c^2 m_{q_2} C_1 C_2} {2^{14} \times 3^2 \times 5 \pi^3} (e_ {q_ 1} + 
   2 e_ {q_ {12}} + e_ {q_ 2}) I[0, 1]\, I_2[h_\gamma]\nonumber\\
  %         \end{align}
%\begin{align}
   &
   +\frac {C_1 f_ {3 \gamma}} {2 ^{23} \times 3^5 \times 5  \pi^3} (4 e_ {q_ {1}} + 8 e_ {q_ {12}} + 
   7 e_ {q_ {2}}) (40 m_c^2 I[0, 2] + 9 I[0, 3]) I_2[\psi^a]\nonumber\\
   &
   -\frac {C_2} {2 ^{20} \times 3 \times 5  \pi^3} (e_ {q_ 2} m_ {q_ {1}} + 
   e_ {q_ {12}} m_ {q_ {12}} + 
   2 e_ {q_ 1} m_ {q_ {2}}) I_ 1[\mathcal S]\, I[0, 4]\nonumber\\
   &
   -\frac {f_{3\gamma} } {2 ^{22} \times 3^3 \times 5^2  \pi^3} (2 e_ {q_ 1} - 2 e_ {q_ {12}} - 
   e_ {q_ 2}) \big(200 m_c^2 I[0, 4] + 27 I[0, 5]\big) I_ 1[\mathcal V],
   \label{app2}
\end{align}
where $C_1 =\langle g_s^2 G^2\rangle$ is gluon condensate and $C_2 =\langle \bar q q \rangle$ is the light-quark condensate.  
The $I[n,m]$, $I_1[\mathcal{F}]$, and~$I_2[\mathcal{F}]$ functions are listed as follows:
\begin{align}
 I[n,m]&= \int_{\mathcal M}^{\rm{s_0}} ds ~ e^{-s/\rm{M^2}}~
 s^n\,(s-\mathcal M)^m,\nonumber\\
 I_1[\mathcal{F}]&=\int D_{\alpha_i} \int_0^1 dv~ \mathcal{F}(\alpha_{\bar q},\alpha_q,\alpha_g)
 \delta'(\alpha_ q +\bar v \alpha_g-u_0),\nonumber\\
 I_2[\mathcal{F}]&=\int_0^1 du~ \mathcal{F}(u),%\nonumber
 \end{align}
 where  $\mathcal M = (m_c+m_b)^2$, and $\mathcal{F}$ denotes the relevant DAs.

\end{widetext}

 Finally, it should be noted that the derivation of Eq. (\ref{edmn04}) is predicated on the assumption that the physical side of the LCSR may be adequately approximated by a single pole.  In the case of the multi-quark states, it is necessary to verify the aforementioned approximation by means of supplementary arguments.   This is due to the fact that the physical representation of the relevant sum rules is also influenced by intermediate two-meson states (for details see for instance~\cite{Weinberg:2013cfa,Lucha:2021mwx,Kondo:2004cr,Lucha:2019pmp}). 
Therefore, it is imperative that the two-meson intermediate effects be taken into consideration when attempting to extract the parameters associated with multi-quark exotic states. 
These terms can be either subtracted from the sum rules or included in the parameters of the pole term. The first approach was primarily utilized in the investigation of pentaquarks~\cite{Sarac:2005fn,Wang:2019hyc,Lee:2004xk}, whereas the second method was employed to study tetraquarks~\cite{Wang:2015nwa,Agaev:2018vag,Sundu:2018nxt,Albuquerque:2021tqd,Albuquerque:2020hio,Wang:2020iqt,Wang:2019igl,Wang:2020cme}.
 In this context, it is necessary to modify the quark propagator in accordance with the following equation: 
\begin{align}
\frac{1}{m^{2}-p^{2}} \longrightarrow  \frac{1}{m^{2}-p^{2}-i\sqrt{p^{2}}\,\Gamma (p)%
},  \label{eq:Modif}
\end{align}%
where $\Gamma (p)$ is the finite width of the multi-quark states generated by the intermediate two-meson contributions.  When these effects are properly taken into account in the sum rules, their contribution to the physical parameters is shown to be roughly less than $5\%$~(see Refs. \cite{Wang:2015nwa,Agaev:2018vag,Sundu:2018nxt,Albuquerque:2021tqd,Albuquerque:2020hio,Wang:2020iqt,Wang:2019igl,Wang:2020cme}), and these contributions do not exceed the inherent limitations of the sum rules computations.   It may be reasonably assumed, therefore, that the results of the magnetic moments will remain unperturbed by the aforementioned effects. Consequently, the contributions resulting from these effects will also remain within the bounds of uncertainty associated with the results in question.  As a result, the contributions of the two-meson intermediate states in the hadronic representation of the correlation function can be safely neglected, and the zero-width single-pole approximation can be employed instead.

\section{Numerical illustrations}\label{numerical}

To obtain the numerical results of the QCD sum rules for magnetic moments, some input parameters are required.  These input parameters are provided in Table  \ref{inputparameter}.   The photon's distribution amplitudes are required to move on to numerical calculations. For completeness, the expressions for the photon DAs and the numerical values of the input parameters that enter these expressions are given %in the Appendix. 
Ref.~\cite{Ball:2002ps}.
%
%\begin{widetext}
 
 \begin{table}[htp]
	\addtolength{\tabcolsep}{10pt}
	\caption{Parameters employed as inputs in the calculations~\cite{Workman:2022ynf,Ioffe:2005ym, Matheus:2006xi,Narison:2018nbv, Agaev:2019kkz,Agaev:2020mqq}.}
	\label{inputparameter}
	%	\begin{center}
%	\begin{ruledtabular}
		%\scalebox{1.0}{
\begin{tabular}{l|c|ccccc}
               \hline\hline
 %               \\
Parameter & Value&Unit \\
 %\\
                                        \hline\hline
$m_u$&$ 2.16 ^{+0.49}_{-0.26}$&MeV       \\
$m_d$&$ 4.67^{+0.48}_{-0.17}$&MeV         \\
$m_c$&$ 1.27 \pm 0.02$&GeV                 \\
$m_b$&$ 4.18^{+0.03}_{-0.02}$&GeV                    \\
$m^{J_\mu^1}_{T_{bc}}$&$  7105 \pm 155 $&MeV                       \\
$m^{J_\mu^2}_{T_{bc}}$&$  7050 \pm 125 $&MeV                         \\
$\langle \bar qq\rangle $&$ (-0.24 \pm 0.01)^3 $&\,\,GeV$^3$                        \\
%$m_0^{2} $&$ 0.8 \pm 0.1 $&\,\,GeV$^2$                        \\
$ \langle g_s^2G^2\rangle  $&$ 0.48 \pm 0.14 $&\,\,GeV$^4$                        \\
%$f_{3\gamma} $&$ -0.0039 $&\,\,GeV$^2$     \\
$f^{J_\mu^1}_{T_{bc}}$&$  (1.0 \pm 0.2) \times 10^{-2}   $&\,\,GeV$^4$                        \\
$f^{J_\mu^2}_{T_{bc}}$&$  (8.3 \pm 1.3) \times 10^{-3}   $&\,\,GeV$^4$                                                                  \\
                                      \hline\hline
 \end{tabular}
%}
%\end{center}
%\end{ruledtabular}
\end{table}

%\end{widetext}

\begin{widetext}
 
\begin{figure}[htp]
\centering
  \subfloat[]{\includegraphics[width=0.4\textwidth]{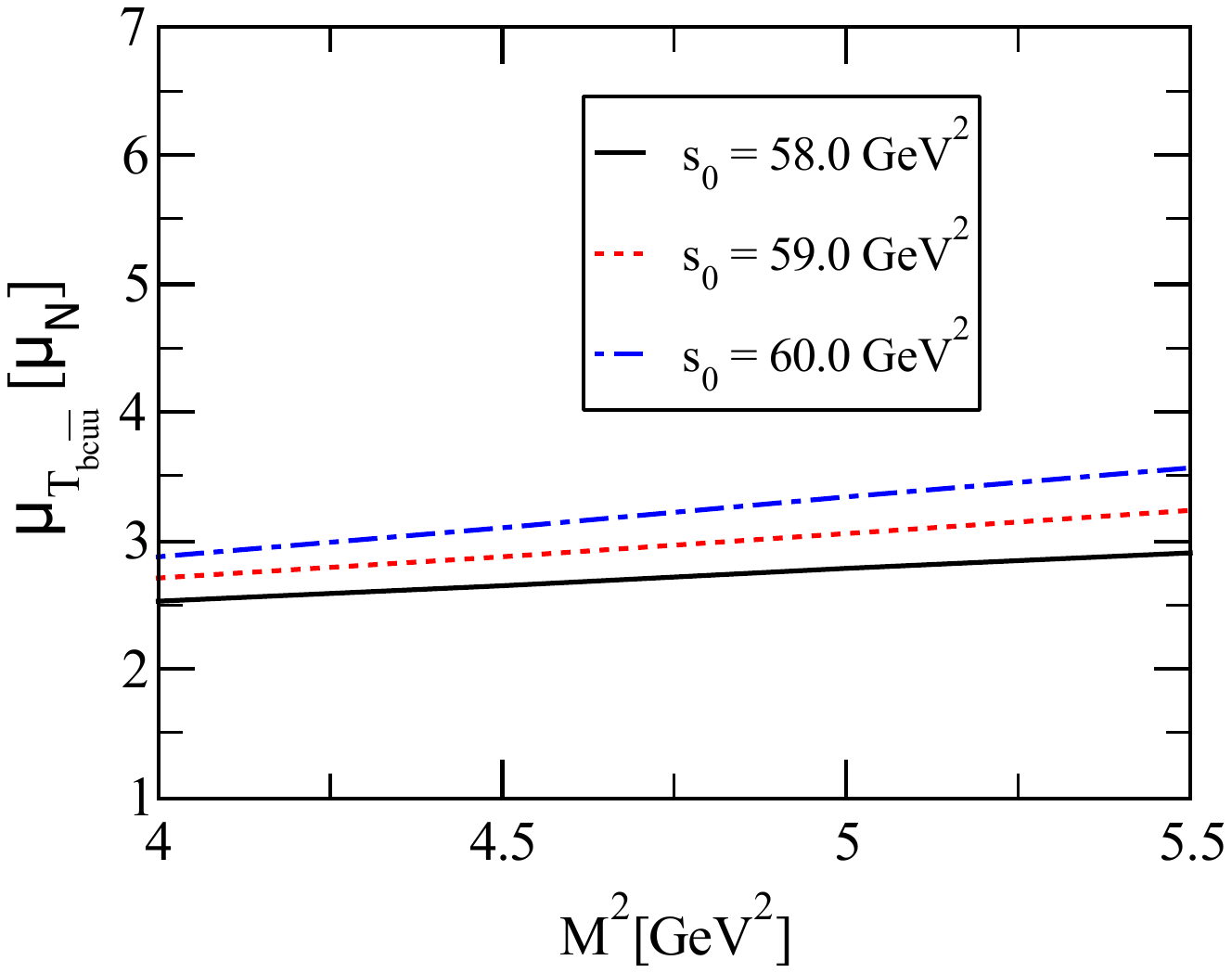}} ~~~~~~~~
   \subfloat[]{\includegraphics[width=0.4\textwidth]{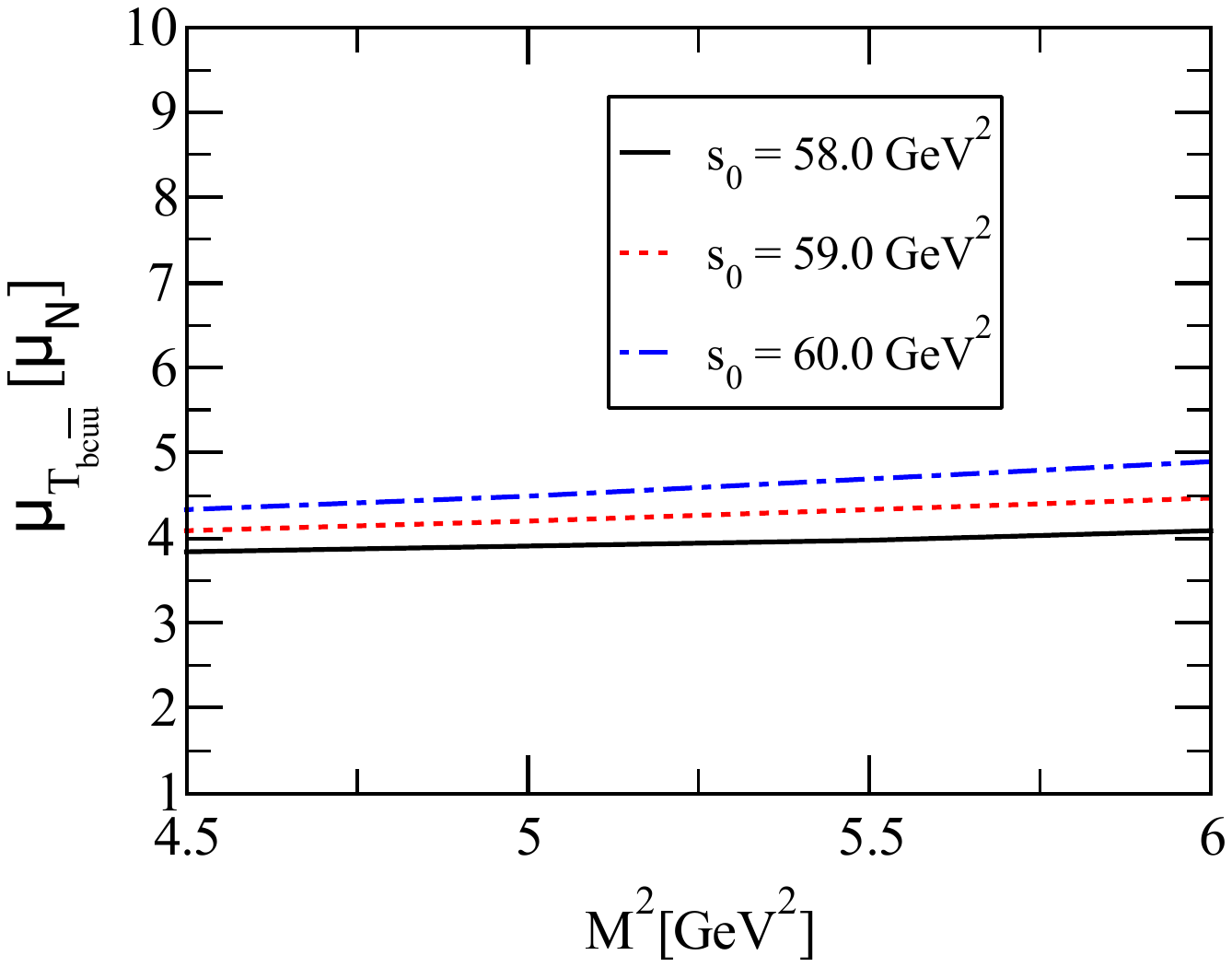}} \\
  \subfloat[]{\includegraphics[width=0.4\textwidth]{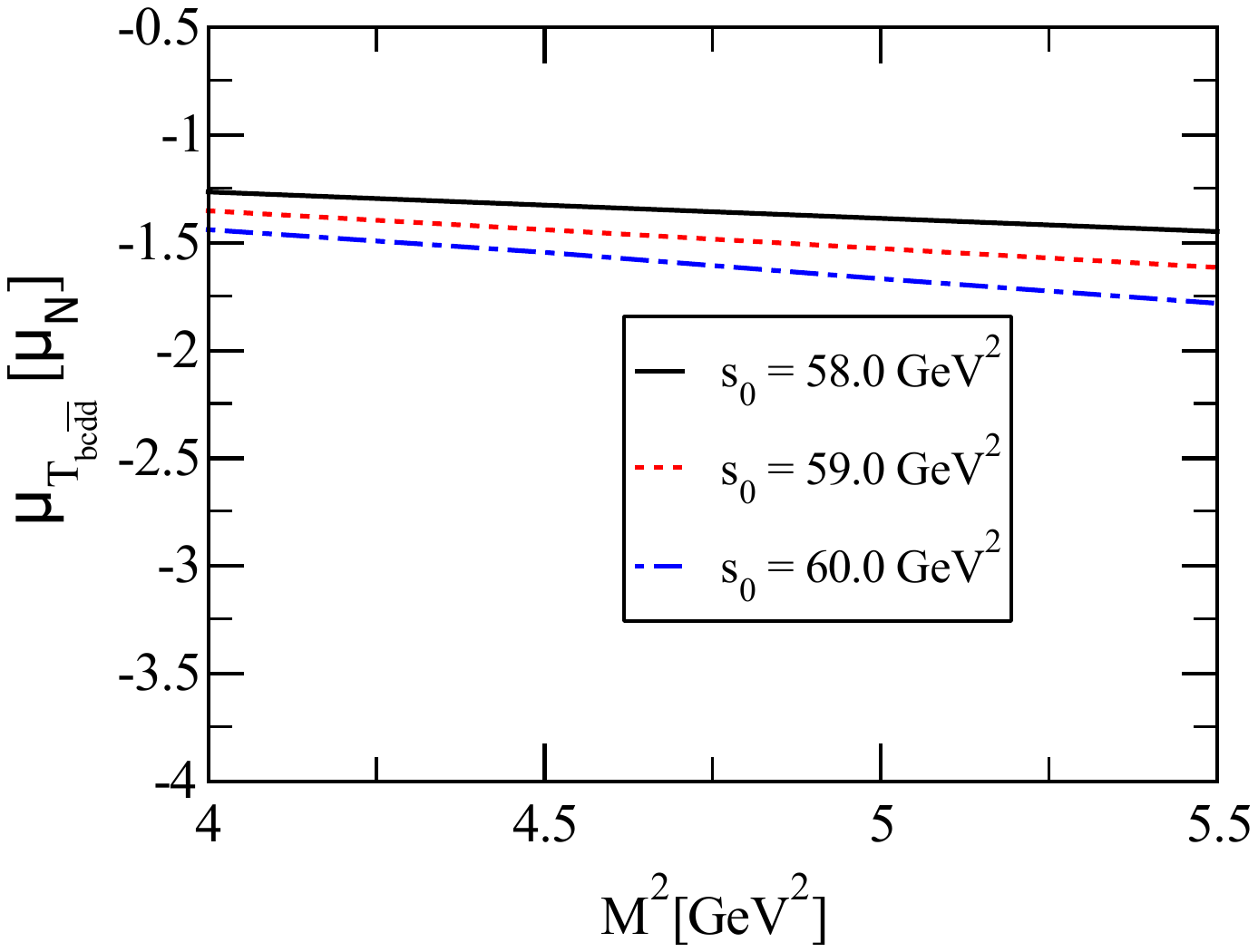}} ~~~~~~~~
   \subfloat[]{\includegraphics[width=0.4\textwidth]{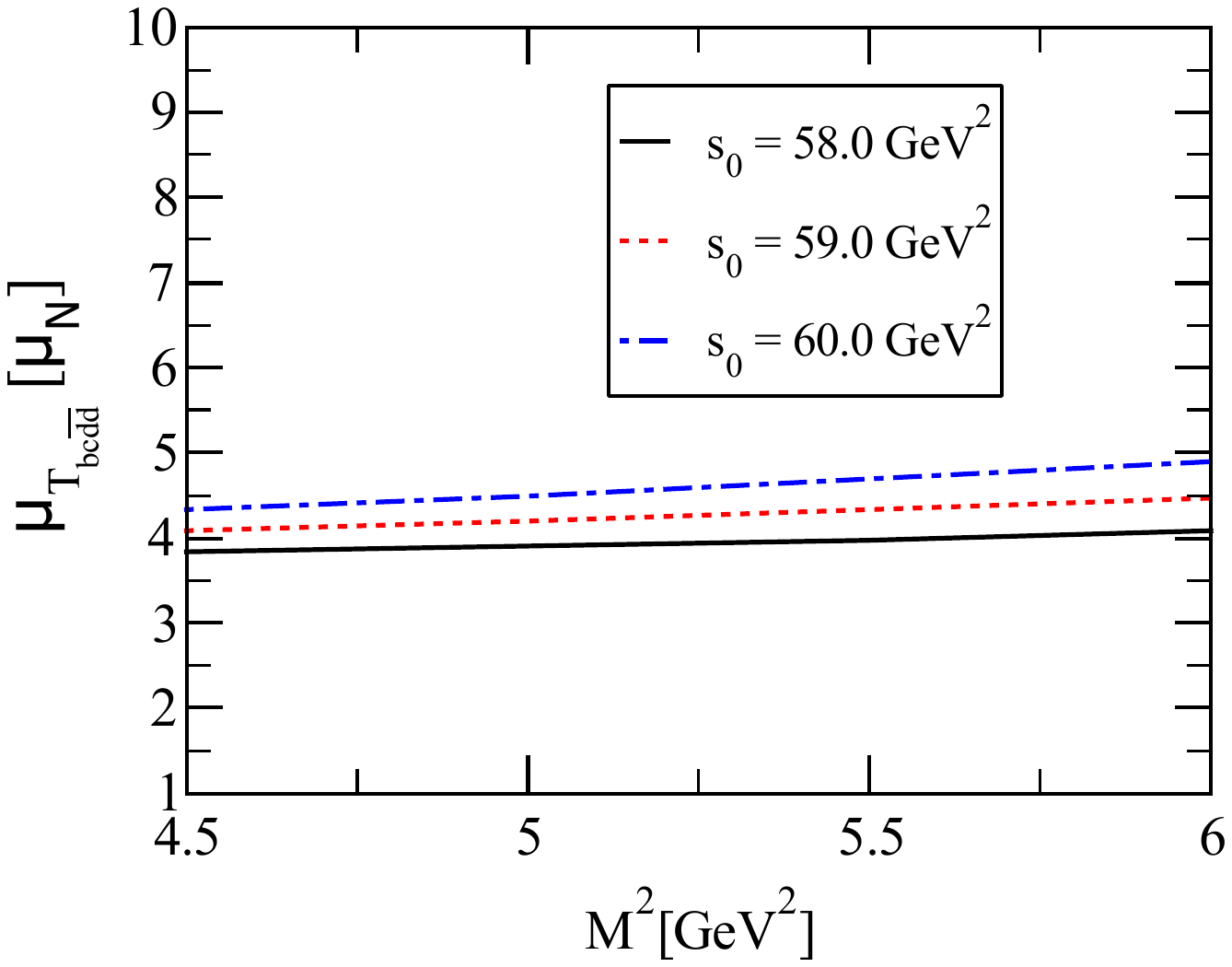} }\\
  \subfloat[]{ \includegraphics[width=0.4\textwidth]{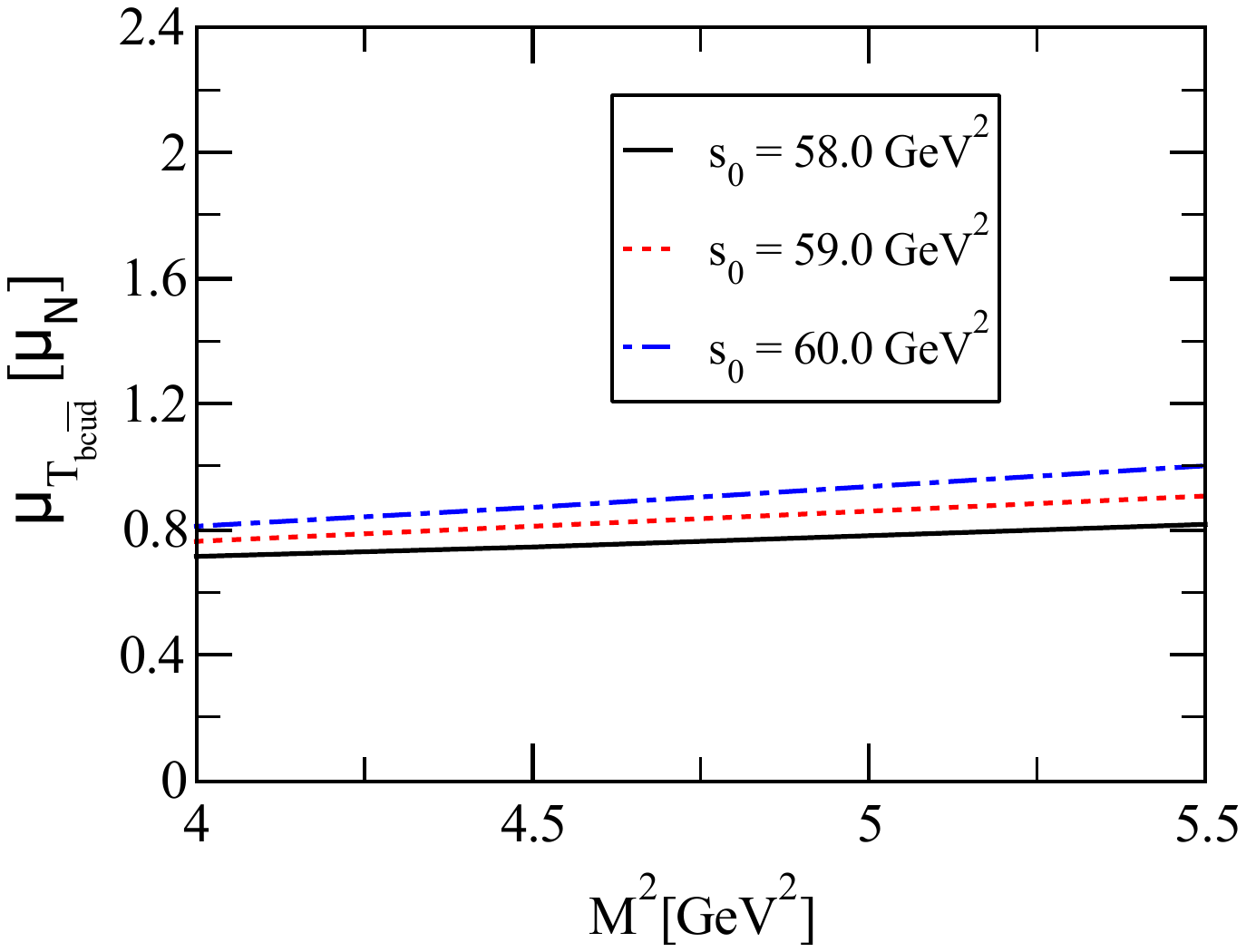}} ~~~~~~~~
  \subfloat[]{\includegraphics[width=0.4\textwidth]{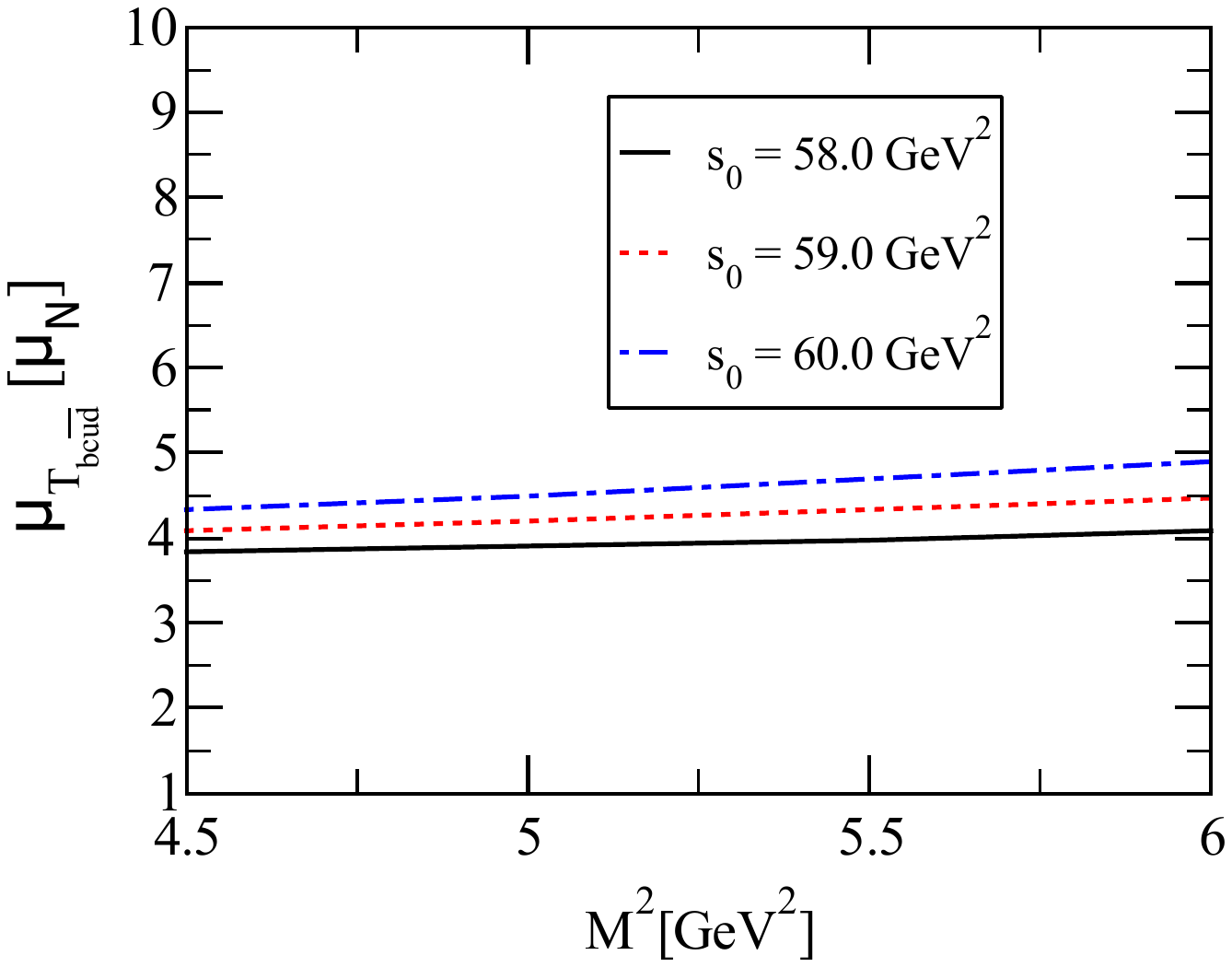}}\\
  \caption{ Variation of magnetic moments $\mu_{T_{bc}}$ as a function of the $\rm{M^2}$ at different values of $\rm{s_0}$; (a), (c)  and (e) for $J^1_\mu$ current, and; (b), (d) and (f) for $J^2_\mu$ current.}
 \label{figMsq}
  \end{figure}

   \end{widetext}

There are two additional parameters, the Borel parameter $\rm{M^2}$ and the threshold parameter $\rm{s_0}$, in addition to those listed in Table \ref{inputparameter}. They are obtained from the analysis of the results based on the standard criteria of the QCD sum rule procedure. These procedures include weak dependence of the results on additional parameters, pole dominance (PC), and convergence of the OPE (CVG). 
It is necessary to impose the following conditions in order to meet these criteria, 
\begin{align}
 \mbox{PC} &=\frac{\rho_i (\rm{M^2},\rm{s_0})}{\rho_i (\rm{M^2},\infty)} \geq  30\%,
 \\
 \nonumber\\
 \mbox{CVG} (\rm{M^2}, \rm{s_0}) &=\frac{\rho_i^{\rm{Dim 7}} (\rm{M^2},\rm{s_0})}{\rho_i (\rm{M^2},\rm{s_0})} \leq 5\%,
 \end{align}
 where $\rho_i^{\rm{Dim 7}} (\rm{M^2},\rm{s_0})$ represents the last term in the OPE  of $\rho_i (\rm{M^2},\rm{s_0})$. 
To guarantee that all requirements are met, numerical analysis is executed and the working regions of the parameters $\rm{M^2}$ and $\rm{s_0}$ are given in Table \ref{table}. The obtained values for the PC and CVG are also presented in the same table for completeness.  
After determining the intervals of these additional parameters, we check the dependence of our results on them and show their behavior in Figure \ref{figMsq}. As can be seen from this figure, these expectations are satisfactorily fulfilled for the chosen working intervals.

Finally, we are ready to estimate the numerical values of the corresponding magnetic moments of the $T_{bc}$ states after setting all the relevant parameters. Predicted magnetic moment results are given in Table \ref{table}. The errors in the predictions, due to errors in the input parameters and uncertainty in the estimates of the working intervals of the ancillary parameters, are shown in Table \ref{table} as well. As we can see from Table \ref{table}, the different interpolating currents that are used to probe the tetraquarks with the same quark content lead to very different results for their magnetic moments. This can be translated into more than one tetraquark state with the same quantum number and quark content yet possessing different magnetic moments. As pointed out earlier, all these interpolating currents have the same quantum numbers, hence leading to almost degenerate masses for these tetraquarks \cite{Agaev:2019kkz,Agaev:2020mqq}. Nevertheless, as can be seen, the results obtained for magnetic moments are quite responsive to the diquark-antidiquark structure and nature of the states under investigation.  In general, it is expected that a change in the basis of the hadrons should not affect the results. However, it is possible that this expectation does not hold true in the case of magnetic moments. This is because magnetic moments are directly related to the inner organization of the hadrons under study. In the language of magnetic moments, changing the basis of the hadron also changes its internal structure, potentially leading to significant changes in the results. Different interpolating currents have been utilized for tetra- and pentaquarks to extract their magnetic moment in~\cite{Azizi:2023gzv, Ozdem:2024rqx, Ozdem:2022iqk}. The results showed that the magnetic moments varied significantly depending on the diquark structures employed.  Consequently, selecting different interpolating currents that may couple to the same states, or altering the isospin and charge basis of the states being studied, can result in varying magnetic moments. We hope that the predictions of the present study can be helpful in our understanding of the bottom-charm tetraquarks quantum numbers and inner structure in experimental measurements.  The consistency of our predictions can be checked by comparing the magnetic moment results obtained in this study with those obtained using different theoretical approaches. It will also be interesting to characterize the branching ratios of the different decay modes and decay channels of the $T_{bc}$ tetraquarks, together with the spectroscopic/decay parameters and magnetic moments.
 \begin{widetext}
 
 \begin{table}[htp]
	\addtolength{\tabcolsep}{10pt}
	\caption{Extracted magnetic  moments  of the $T_{bc}$ states for both $J_\mu^1$ and $J_\mu^2$ interpolating currents with $J^P = 1^+$.}
	%\scriptsize
	\label{table}
	\begin{center}
\begin{tabular}{lccccccc}
	   \hline\hline
	   \\
	   Current&  $\mu_{T_{{bc \bar u \bar u}}}$[$\mu_N$]	& $\mu_{T_{{bc \bar d \bar d}}}$[$\mu_N$]& $\mu_{T_{{bc \bar u \bar d}}}$[$\mu_N$]	& $\rm{M^2}[\rm{GeV}^2]$& $\rm{s_0}[\rm{GeV}^2]$& PC[$\%$] & CVG[$\%$]	   \\
	   \\
	  %\\
	   \hline\hline
	  \\
	   $J_\mu^1$&   $ 2.97^{+0.59}_{-0.44}$     &$-1.48^{+0.30}_{-0.22}$ &	$ 0.84^{+0.16}_{-0.13}$& [4.0, 5.5]& [58, 60]& [62, 31]& 3.14\\
	 %  \\
	% \\
	   %\hline\hline
	   \\
	   $J_\mu^2$& $ 4.27^{+0.62}_{-0.43}$     &$4.31^{+0.62}_{-0.44}$ &	$ 4.29^{+0.61}_{-0.44}$& [4.5, 6.0]& [58, 60]& [60, 32]& 2.85
	   \\
	   \\
	   \hline\hline
\end{tabular}
\end{center}
\end{table}

 \end{widetext}
 
A final comment would be to determine the individual quark sector contributions to the magnetic moments to better understand the underlying quark-gluon dynamics. This can be attained by choosing the relevant charge factors $e_b$, $e_c$, and $e_q$. 
 Our analyses show that the magnetic moments of the $T_{bc}$ states that are characterized by the $J_\mu^1$ interpolating current are governed by the light-diquarks, while those of the $T_{bc}$ states that are characterized by the $J_\mu^2$ interpolating current are governed by the heavy-diquarks.
In order to gain a deeper insight into the underlying reasons for this phenomenon, it is necessary to analysis the structure of the interpolating currents employed in the study. In the case of the $J_\mu^1$ interpolating current, the heavy-diquark component is constructed on the basis of the scalar diquark structure, while the light-diquark component is constructed on the basis of the axial-vector diquark structure. In the case of the $J_\mu^2$ interpolating current, the heavy-diquark component is constructed on the basis of the axial-vector diquark structure, while the light-diquark component is constructed on the basis of the scalar diquark structure. As is well known, the contribution of scalar or pseudoscalar structures to magnetic moments is almost negligible. Therefore, it is expected that the axial-vector diquark component will make the dominant contribution to our analysis. This fact leads to the conclusion that the dominant contribution to the results obtained using the $J_\mu^1$ interpolating current comes from light quarks, while for the $J_\mu^2$ interpolating current the dominant contribution comes from heavy quarks. As the type of heavy quarks utilized in this study remains unchanged, the outcomes yielded by the $J_\mu^2$ interpolating current are anticipated to exhibit minimal variation. Conversely, the outcomes derived from the $J_\mu^1$ interpolating current are expected to change in accordance with the type of light quarks employed.

\section{Final remarks}\label{sum}

In this study, the magnetic moments of possible axial-vector $T_{bc}$ tetraquarks are obtained with the help of light-cone QCD sum rules. For this purpose, we assume that these states are represented as a diquark-antidiquark picture with different structures and interpolating currents. The magnetic moment results derived using different diquark-antidiquark configurations differ substantially from each other. This can be translated into more than one tetraquark state with the same quantum number and quark content yet possessing different magnetic moments. From the numerical results obtained, we have concluded that the magnetic moments of the $T_{bc}$ states can project their inner structure, which can be used for their quantum numbers and quark-gluon organization. For the sake of completeness, we also analyze the contribution of the individual quarks to the magnetic moments. We observe that the magnetic moments of the $T_{bc}$ states that are characterized by the $J_\mu^1$ interpolating current are governed by the light-diquarks, while those of the $T_{bc}$ states that are characterized by the $J_\mu^2$ interpolating current are governed by the heavy-diquarks. 
We hope that our predictions of the magnetic moments of the $T_{bc}$ tetraquarks, together with the results of other theoretical investigations of the spectroscopic parameters and decay widths of these interesting tetraquarks, may be valuable in the search for these states in future experiments and in unraveling the internal structure of these tetraquarks.

% \textbf{Data Availability Statement:} This manuscript has no associated data or the data will not be deposited. [Authors’ comment: This is a theoretical research work, so no additional data are associated with this work.]

\bibliography{TbcMM.bib}
\bibliographystyle{elsarticle-num}

\end{document}